\documentstyle[aps,twocolumn,twoside]{revtex}

\author{ D.V. Savin \thanks{e-mail: savin@inp.nsk.su} }
\address{ Budker Institute of Nuclear Physics, 630090 Novosibirsk, Russia }
\title{ Random band matrix approach to chaotic scattering:  \\
        the average $S$-matrix and its pole distribution. }
\date{ November 28, 1995 }

\begin{document}
\draft
\maketitle

\begin{abstract}
 Random band matrices relevant for open chaotic systems are introduced
 and studied. The scattering model based on such matrices may serve for the
 description of preequilibrium chaotic scattering. In the limit of a large
 number of open channels we calculate the average $S$-matrix and $S$-matrix's
 pole distribution which are found to reduce to those of the full matrix
 (GOE) case under proper renormalization of the energy scale and strength of
 coupling to the continuum.
\end{abstract}
\pacs{ }

\narrowtext

{\bf 1.} It is now generally acknowledged that the Random Matrix Theory (RMT)
provides an adequate tool for the description of statistical properties of
chaotic quantum systems \cite{Mehta}--\cite{Brody}.  The most of works in
the RMT deals with the Hermitian matrices which are, strictly speaking,
appropriate for the description of closed systems.  Meanwhile, any exited
state of a quantum system has a finite lifetime, decaying eventually into
open channels. It is well known \cite{MW-69}--\cite{SZ-89} that the openness
of a system can be incorporated into consideration by means of the effective
nonhermitian Hamiltonian
\begin{equation}\label{ham}
 {\cal H}_{nm} = H_{nm} - i \sum_{c\,(open) } V^c_n V^c_m\  .
\end{equation}
Due to coupling to the continuum, the effective Hamiltonian (\ref{ham})
acquires, apart from the intrinsic part $H$, the antihermitian part
which consists of the sum of products of the transitions amplitudes $V^c_n$
between $N$ internal $(|n\rangle)$ and $M$ open channel $(|c\rangle)$ states.
For the T-invariant theory these amplitudes as well as $H$
can be chosen to be real, the matrix ${\cal H}$ being symmetric.
The eigenvalues ${\cal E}_n=E_n-\frac{i}{2}\Gamma_n$ of ${\cal H}$ are
the complex energy levels of an unstable system, with $E_n$
and $\Gamma_n$ being, respectively, the energy and width of the $n$-th level.

  Assuming the intrinsic dynamics to be chaotic, the hermitian part of
${\cal H}$ is usually supposed to belong to the Gaussian Orthogonal Ensemble
(GOE) of random matrices. As to the coupling amplitudes, they are considered
to be either fixed \cite{W-84,VWZ-85} or random \cite{SZ-89}.  Resulting
generalization of the RMT on the unstable systems proved
\cite{W-84}--\cite{LSSS-95b} to be useful for the description of
statistical properties of quantum chaotic scattering.
The resonance part of the scattering matrix is represented (see, e.g.
\cite{MW-69}) as
\begin{equation}\label{S}
 S_{ab}(E) = \delta_{ab} -
             2i \sum_{nm} V^a_n[(E-{\cal H})^{-1}]_{nm}V^b_m\ .
\end{equation}
in terms of the effective Hamiltonian (\ref{ham}) which describes the
evolution of the unstable system formed at the intermediate stage of the
collision. Therefore, fluctuations in scattering reflect statistical
properties of the complex energy levels of this intermediate system.

  Recently, the considerable progress has been achieved in the study of
dynamical and statistical properties of $S$-matrix's poles and their
connection to those of scattering \cite{SZ-89} -\cite{LSSS-95b}. The limit
of very large number of equivalent channels has a special interest since
it can be related to the semiclassical approach \cite{LW-91,LSSS-95a}.
The  noteworthy property of the pole distribution was found
\cite{HILSS-92,LSSS-95a} in this limit: there always exists a finite gap
between the upper edge of the distribution of complex energies of resonances
and the real energy axis. This gap turns out to be the important
characteristic of local fluctuations in chaotic scattering
\cite{LSSS-95a,LSSS-95b}.

 The Gaussian distribution of the matrix elements of the internal
Hamiltonian $H$ implies the invariance with respect to the choice of the
intrinsic basis. However, the mean-field basis seems to play an exceptional
role \cite{Z-93} in many-body chaotic systems. The realistic Hamiltonian
matrix has a banded structure in such a representation
\cite{FGGK-94,ZHB-95}. In fact, Wigner was the first who considered band
matrices in connection with properties of (stable) complicated systems
\cite{W-55}.  He assumed the nonzero matrix elements to be Gaussian random
variables.  Now, the theory of random band matrices (RBM) attracts a great
interest and is claimed to be relevant for studying quantum chaos (for review
see \cite{CC-95}).  The most of known analytical results for RBM has been
recently obtained by Fyodorov and Mirlin \cite{FM-91,FM-95} who studied RBM
in the context of the localization, having reduced the RBM problem to the
Efetov's supersymmetric $\sigma$-model \cite{E-83} with the diffusion
constant proportional to the square of the bandwidth.

   The purpose of this Letter is to extend the RBM approach to the
consideration of open chaotic systems described by the effective Hamiltonian
(\ref{ham}). The scattering model based on such random matrices can be
related to preequilibrium chaotic scattering in a sense intermediate between
fully chaotic scattering described by the GOE model and multistep
compound reactions \cite{NVWY-86}.  In the "pure" GOE models all internal
degrees of freedom being uniformly involved, the intermediate unstable system
has already attained its complete thermodynamical equilibrium before a
decay takes place.  The band structure of $H$ leads to the localization of
intrinsic wave functions, an intermediate decaying state concluding only a
part of the degrees of freedom. Therefore, in addition to the decay time, the
new
diffusion time scale appears in the problem which characterizes the internal
relaxation time. In this Letter we calculate the average $S$-matrix and
$S$-matrix's pole distribution in the limit of very large number of open
equivalent channels.

{\bf 2.} We suppose that the hermitian part of (\ref{ham}) belongs to the
Gaussian RBM ensemble defined by
\begin{equation}\label{rbm}
  \langle H_{nm}H_{nm}\rangle = \lambda^2 J_{nm}(1+\delta_{nm}) \ ,
\end{equation}
where the function $J_{nm}\equiv J(|n-m|)$ decreases sufficiently fast
then $|n-m|>b$, with $b\gg 1$ being the effective bandwidth. In the GOE case,
when all matrix elements $J_{nm}$ are equal to $\frac{1}{N}$, $2 \lambda$
determines the radius of Wigner's semicircle low of the eigenvalue
distribution \cite{Mehta}. The transition amplitudes $V^a_n$
are considered as fixed quantities subject to the condition
\cite{W-84,VWZ-85}
\begin{equation}\label{ampl}
   \sum_n V^a_nV^b_n = \gamma \lambda \delta^{ab} \ ,
\end{equation}
which turns out to be enough for excluding direct reactions.
The dimensionless parameter $\gamma$ characterizes  the strength of coupling
between the internal motion and channels.

  We calculate the average $S$-matrix and distribution of complex energy
levels in the limit $M, N\rightarrow\infty$, with $m=M/N<1$ fixed.  It turns
out that the band structure of the hermitean part $H$ does not lead to the
essential complication when evaluating the one-point average characteristics
and all calculations can be done in close analogy with those of the GOE case
\cite{HILSS-92,LSSS-95a}.  Below we present rather short description,
referring for details to \cite{HILSS-92,LSSS-95a}.

{\bf 3.} The calculation of the average $S$-matrix is related with that of
the average Green function
\begin{equation}\label{green}
 \langle {\cal G}(z) \rangle =
 \Bigl\langle \left(\frac{1}{z-{\cal H}}\right) \Bigr\rangle \, ,
 \ \  z=E+i0 \ ,
\end{equation}
which governs the evolution of the intermediate unstable
chaotic system.  It is convenient to use the following representation for
$\langle{\cal G}_{nm}(z)\rangle$ \cite{W-84}
\begin{eqnarray}\label{green-z}
 & & \langle {\cal G}_{nm}(z) \rangle =
 2\frac{\partial}{\partial I_{nm}}\langle\ln Z(z,I)\rangle\Big|_{I=0} \ , \\
 & & Z(z,I)= \det(z - H + i VV^T - I)^{-\frac{1}{2}} \ .  \nonumber
\end{eqnarray}

 To carry out averaging in (\ref{green-z}) we use the replica method
\cite{EA-75,W-84}.  The generating function $Z(z,I)$ can be represented as a
multivariable Gaussian integral that makes averaging over the Gaussian RBM
ensemble (\ref{rbm}) trivial. The further integration can be performed
by means of the saddle-point approximation.  The saddle-point solution turns
out to be stable in the replica space and proportional to $\delta_{nn'}$.
Therefore, the replicas decouple  and, as was pointed in \cite{HILSS-92}, it
is enough to calculate $\ln\langle Z(z,I) \rangle$.

One has
\begin{eqnarray}\label{z-aver}
 \langle Z(z,I)\rangle \cong \int\!& d [\phi]
  & \,\exp\Bigr\{ \sum_{n,m=1}^N \bigr[ - \frac{\lambda^2 }{4}
   J_{nm}\phi^2_n \phi^2_m \nonumber \\
  & + & \frac{i}{2}\phi_n(z+i VV^T-I)_{nm} \phi_m \bigl] \Bigl\}  \ ,
\end{eqnarray}
where $d[\phi]$ means the product of differentials and above equality is
valid to irrelevant constant. To make the integration over $\phi$ doable,
we introduce, following \cite{HILSS-92,LSSS-95a}, new variables
$\sigma_n=\lambda\phi_n^2$ with the help of $\delta$-functions defined by
their Fourier representation
\begin{equation}\label{delta}
  \delta(\sigma_n-\lambda\phi_n^2) = \frac{1}{\pi}\int\!d\hat\sigma_n \exp
  \{ - \frac{i}{2}(\sigma_n\hat\sigma_n-\lambda\phi_n\hat\sigma_n\phi_n) \}
\end{equation}
After the Gaussian integration over $\phi$ being done, the subsequent
integration can be performed in the saddle-point approximation justified by
two large parameters $N,b \gg 1$. We find after some algebra that the average
Green function is determined in the following way
\begin{equation}\label{green-l}
 \langle {\cal G}_{nm}(z) \rangle = \bigl[(z + \lambda\hat\sigma_{s.p.}
 + i VV^T )^{-1}\bigr]_{nm}
\end{equation}
by the solution of the saddle-point equations
(with respect to $\sigma$ and $\hat\sigma$) for the "Lagrangian" ${\cal L}$
\begin{eqnarray}\label{lagr}
 {\cal L} &=& - \frac{1}{4}\sum_{nm}J_{nm}\sigma_n\sigma_m  - \sum_n
 (\frac{i}{2}\sigma_n\hat\sigma_n + \ln(z+\lambda\hat\sigma_n)) \nonumber \\
 &\ & - \frac{1}{2}\mbox{tr}_{c}\ln [1+i V^T(z+\lambda\hat\sigma)^{-1}V] \ ,
\end{eqnarray}
where the diagonal matrix
$\hat\sigma=\mbox{diag}(\hat\sigma_1,\ldots,\hat\sigma_N)$. Note that the
trace in the last term in (\ref{lagr}) runs over the channel
($M$-dimensional) space.

   Similar to the pure RBM case \cite{FM-91}, the saddle point equations
possess the translation invariant (independent of $n$) solution provided that
the relation (\ref{ampl}) is fulfilled. Going along the same line as in
\cite{HILSS-92,LSSS-95a}, we arrive at
\begin{equation}\label{green-aver}
 \langle{\cal G}_{nm}(z)\rangle =
  \frac{ \delta_{nm}(z - J_0\lambda^2 g(z) + i\gamma\lambda)
  - i \sum_c V^c_nV^c_m }{ (z-J_0\lambda^2 g(z) )
  ( z-J_0\lambda^2 g(z)+i\gamma\lambda) } \ .
\end{equation}
Here the notation $J_0\equiv\sum_r J(|r|)$ is introduced.
Due to the translation invariance mentioned above the average Green function
depends on the only combination of transition amplitudes. This fact makes
the average $S$-matrix be diagonal and  equal to
\begin{equation}\label{s-aver}
 \langle S^{ab}(z)\rangle = \delta^{ab}
   \frac{z - J_0\lambda^2 g(z) - i\gamma\lambda}{
         z - J_0\lambda^2 g(z) + i\gamma\lambda    } \ .
\end{equation}
The function $g(z)$ denotes the trace of the average Green function which is
found to satisfy the cubic equation
\begin{equation}\label{cubic}
 g(z)\bigl(J_0\lambda^2 g(z) - z \bigr) + 1
  + \frac{i m\gamma\lambda}{J_0\lambda^2 g(z) - z - i\gamma\lambda} = 0 \ ,
\end{equation}
where the (unique) solution with a negative imaginary part has to be chosen
\cite{LSSS-95a}.

  For a closed system ($\gamma=0$), this cubic equation is reduced to the
quadratic one which determines the density of the RBM eigenvalues to be
given by Wigner's semicircle law \cite{KLH-91} with the half-radius
\begin{equation}\label{scale1}
 \widetilde \lambda = \lambda\sqrt{J_0} \ ,
\end{equation}
renormalized by the factor $\sqrt{J_0}$ (this factor reduces to unity in the
case of the GOE). For an open system, additional rescaling of the coupling
constant
\begin{equation}\label{scale2}
 \widetilde \gamma = \frac{\gamma}{\sqrt{J_0}},
\end{equation}
reduces eqs.(\ref{green-aver})--(\ref{cubic}) to corresponding ones
\cite{SZ-92,LSSS-95a} for the full matrix case.

{\bf 4.} The calculation carried out above is valid only for the upper half
of the complex plane where the Green function is analytical, all $S$-matrix
singularities being located into the lower part (see eq.(\ref{S})). The
analyticity is broken where the density of complex energies differs from
zero. A special regularization procedure of the pole singularities has been
proposed in \cite{HILSS-92} to calculate the average $S$-matrix's pole
distribution in the lower halfplane of the complex variable $z=x+iy$.  Basing
on a convenient electrostatic analogy \cite{SZ-89,HILSS-92}, the distribution
of complex levels is considered to be the source of the two-dimensional
electrostatic field with the potential
\begin{equation}\label{pot}
 \Phi(x,y) = \frac{1}{N} \langle \ln\det \bigl\{
  (z^*-{\cal H}^{\dagger})(z-{\cal H})+\delta \bigr\}^{-1} \rangle \ .
\end{equation}
The limit $\delta\rightarrow 0^+$ should be taken at the very end of
calculations. Applying the two-dimensional Laplacian to $\Phi$, one gets the
average density of complex levels ("charges")
\begin{equation}\label{dens}
 4\pi\rho(x,y) = - \Delta  \Phi(x,y)  \ .
\end{equation}

  The replica trick can be  used again for performing the ensemble averaging.
The infinitesimal positive $\delta$ makes the matrix in the rhs of
eq.(\ref{pot}) positive definite  for any $z$. Therefore, the determinant may
be represented as the Gaussian integral over a complex $N$-vector  $\psi(1)$.
To make the ensemble averaging possible, the product
$(z^*-{\cal H}^{\dagger})(z-{\cal H})$ is decoupled \cite{HILSS-92,LSSS-95a}
with the help of the complex Habbard-Stratonovich transformation
($N$-vector $\psi(2)$). One finally arrives at the integral representation
\begin{equation}\label{pot-l}
  \exp\{ N\Phi \} \cong \int\!d[\psi]\,\bigl\langle\,
  \exp\{i \psi^{\dagger} {\cal M} \psi' \} \bigr\rangle
\end{equation}
where $2N\times 2N$ matrix ${\cal M}$ is defined as
$$ {\cal M}=\left[ \begin{array}{cc}
    z^{\ast}-{\cal H}^{\dagger} &  i \delta  \\
    i    &       z-{\cal H}  \end{array}  \right] \ ,
$$
introduced $2N$-vectors being $\psi^T=\bigl(\psi(1)^T,\psi(2)^T\bigr)$
and $\psi'^T=\bigl(\psi(2)^T,\psi(1)^T\bigr)$.

Performing the same steps as in \cite{HILSS-92,LSSS-95a}, we find
the potential $N\Phi$ to be determined by the saddle-point value of
the "Lagrangian"
\begin{eqnarray}\label{l-dens}
&& {\cal L} =
   - \frac{1}{2}\sum_{nm} J_{nm}\mbox{tr}_{\alpha}(\sigma_n\sigma_m)
   - i\sum_n \mbox{tr}_{\alpha}(\sigma_n\hat\sigma_n)      \\
&& - \sum_n\mbox{tr}_{\alpha}\ln(z_{\delta}+\lambda\hat\sigma_n) \nonumber
  - \mbox{tr}_{\alpha c} \ln \bigl[1 -
         i V^T(z_{\delta}+\lambda\widehat\Sigma)^{-1}V l\bigr]  \  .
\end{eqnarray}
Each of $N$ matrices $\sigma_n$ in (\ref{l-dens}) has the following structure
$$
 \sigma=\left( \begin{array}{cc} w & u \\  v & w^{\ast} \end{array} \right)
$$
with the real positive $u$ and $v$ whereas $\hat\sigma_n$ stands for its
Fourier counterpart. We have also introduced $2N\times 2N$ block-diagonal
matrices $z_{\delta}$ obtained from ${\cal M}$ by setting there $\cal H$
equal to zero  and $\widehat\Sigma_{\alpha\beta\ nm}=(\hat\sigma_n)_{\alpha
\beta} \delta_{nm}$, the $2M\times 2M$ block-diagonal matrix $l$ is unity
in the channel subspace and equals to $l=\mbox{diag}(1,-1)$ in the replica
subspace, and $V$ is the unit matrix for replica indices and corresponds to
$V^{c}_{n}$ for others. In the saddle-point we have the translation
invariant saddle-point equations
\begin{eqnarray}\label{saddle}
&& \hat\sigma=iJ_0\sigma   \nonumber \\
&& \frac{i\sigma}{\lambda}(z_{\delta} + i\lambda J_0 \sigma ) + 1
   - i\frac{m\gamma\lambda l}{ z_{\delta}
   + i\lambda J_0 \sigma - i\gamma\lambda l} = 0
\end{eqnarray}
which differ from the corresponding equations for the full matrix case only
in appearing the renormalizing factor $J_0$. Therefore, the explicit
solution can be found in our case in the same way as it has been done in
\cite{LSSS-95a}. As a result, one concludes that all poles (charges) lie in
the finite domain of the lower part ($y<0$) of the complex energy plane
defined by the condition $x^2 \leq {\sf x}^2(y)$ with
\begin{equation}\label{bound}
 {\sf x}^2(y) =
  - \frac{4m\gamma\lambda^3 J_0}{y} - \Bigl[ \frac{m\lambda^2 J_0}{y}
  + \frac{1-m}{\gamma\lambda + y}\lambda^2 J_0 - \gamma\lambda \Bigl]^2\ .
\end{equation}
Inside this region the density of complex energy levels is equal to
\begin{equation}\label{dens-aver}
  4\pi\rho(x,y) = \frac{m}{y^2} + \frac{1-m}{(\gamma\lambda+y)^2}
                  - \frac{1}{J_0\lambda^2} \ .
\end{equation}
One can easily see that under rescaling (\ref{scale1}),(\ref{scale2}) the
average complex level distribution reduces again to that \cite{LSSS-95a}
of the full matrix case.

  In conclusion, the band structure of the hermitean part $H$ results only in
the renormalization of the energy scale $\lambda$ (\ref{scale1}) and coupling
constant $\gamma$ (\ref{scale2}) as compared to the GOE model
\cite{LSSS-95a}. In particular, the condition of the "width collectivization"
\cite{SZ-89,SZ-92}, $\widetilde\gamma \sim 1$, implies again  the natural
physical condition of the average partial width being comparable with the
average level spacing \cite{SZ-92}. One should expect nontrivial consequences
of the band structure to appear only while considering the higher correlation
functions.

  I am indebted to V.V. Sokolov for bringing my interest to the subject,
critical reading of the manuscript, and illuminating discussions.  The
partial financial support from the International Science Foundation (grants
RB7000 and RB7300) and INTAS (grant 94-2058) is acknowledged with thanks.


\begin{thebibliography}{99}

\bibitem{Mehta} M.L. Mehta, {\em Random Matrices}, (Academic Press, NY, 1991).

\bibitem{Porter} C.E. Porter, {\em Statistical Theories of Spectra:
 Fluctuations}, (Academic Press, NY, 1965).

\bibitem{Brody} T.A. Brody, J. Flores, J.B. French, P.A. Mello, A. Pandey,
 S.S.M. Wong, Rev. Mod. Phys. {\rm 53} (1981) 385.

\bibitem{MW-69} C. Mahaux and H.A.  Weidenm\"uller, {\em Shell-model
 Approach to Nuclear Reactions}, (North-Holland, Amsterdam, 1969).

\bibitem{W-84} H.A. Weidenm\"uller, Ann. of Phys. {\rm 158} (1984) 120.

\bibitem{VWZ-85} J.J.M. Verbaarschot, H.A. Weidenm\"uller, and
 M.R. Zirnbauer,  Phys. Rep. {\rm 129} (1985) 367.

\bibitem{SZ-89} V.V. Sokolov and V.G.  Zelevinsky, Phys. Lett. {\rm B202}
 (1988) 140; Nucl. Phys. {\rm A504} (1989) 562.

\bibitem{HILSS-92} F. Haake, F.M. Izrailev, N. Lehmann, D. Saher, and
 H.-J. Sommers, Z. Phys. {\rm B88} (1992) 359.

\bibitem{LSSS-95a} N. Lehmann, D.  Saher, V.V. Sokolov, and H.-J. Sommers,
 Nucl. Phys. {\rm A582} (1995) 223.

\bibitem{LSSS-95b} N. Lehmann, D.V. Savin, V.V. Sokolov, and H.-J. Sommers,
 Physica D {\rm 86} (1995) 572.

\bibitem{LW-91} C.H. Lewenkopf and H.A. Weidenm\"uller, Ann. Phys. (N.Y.)
 {\rm 212} (1991) 53.

\bibitem{Z-93} V.G. Zelevinsky, Nucl. Phys. {\rm A555} (1993) 109.

\bibitem{FGGK-94} V.V. Flambaum, A.A. Gribakina, G.F. Gribakin, and
 M.G. Kozlov, Phys. Rev. {\rm A50} (1994) 267.

\bibitem{ZHB-95} V. Zelevinsky, M. Horoi, and B.A. Brown, Phys. Lett.
 {\rm B350} (1995) 141; M. Horoi, V. Zelevinsky, and B.A. Brown,
 Phys. Rev. Lett. {\rm 74} (1995) 5194.

\bibitem{W-55} E. Wigner, Ann. Math. {\rm 62} (1955) 548; ibid {\rm 65}
 (1957) 203.

\bibitem{CC-95} G. Casati and B.V. Chirikov, eds., {\em Quantum Chaos},
 (Cambridge Univ. Press, Cambridge, 1995).

\bibitem{FM-91} Y.V. Fyodorov and A.M. Mirlin, Phys. Rev. Lett. {\rm 67}
 (1991) 2405; ibid {\rm 69} (1992) 1093; ibid {\rm 71} (1993) 412;

\bibitem{FM-95} Y.V. Fyodorov and A.M. Mirlin, Int. J. Mod. Phys. {\rm 8}
 (1994) 3795.

\bibitem{E-83} K.B. Efetov, Advan. in Phys. {\rm 32} (1983) 53.

\bibitem{NVWY-86} H. Nishioka, J.J.M. Verbaarschot, H.A. Weidenm\"uller,
  and S. Yoshida, Ann. Phys. {\rm 172} (1986) 67.

\bibitem{EA-75} S.F. Edwards and P.W. Anderson, J. Phys. F{\rm 5} (1975) 965.

\bibitem{KLH-91} M. Kus, M. Lewenstein, and F. Haake, Phys. Rev. {\rm A44}
 (1991) 2800.

\bibitem{SZ-92} V.V. Sokolov and V.G.  Zelevinsky, Ann. Phys. {\rm 216}
 (1992) 323.

\end{thebibliography}
\end{document}